\documentstyle[aps,epsfig,prl,floats]{revtex}

\newcommand{\be}{\begin{equation}}
\newcommand{\ee}{\end{equation}}
\newcommand{\bea}{\begin{eqnarray}}
\newcommand{\eea}{\end{eqnarray}}
\newcommand{\bdm}{\begin{displaymath}}
\newcommand{\edm}{\end{displaymath}}

\newcommand{\livo}{Li$_2$VOSiO$_4$}
\newcommand{\ligo}{Li$_2$VOGeO$_4$}

\tightenlines
\begin{document}
\draft

\title{A closer look at symmetry breaking in
the collinear phase of the $J_1-J_2$ Heisenberg Model}

\author{
R.R.P.~Singh$^{(a)}$, W. Zheng$^{(b)}$, J. Oitmaa$^{(b)}$, O.P. Sushkov$^{(b)}$,
and C.J.~Hamer$^{(b)}$
 }
\address{
$^{(a)}$ Department of Physics, University of California, Davis, CA
95616\\
$^{(b)}$ School of Physics, University of New South Wales, Sydney NSW
2052, Australia\\
}

\twocolumn[\hsize\textwidth\columnwidth\hsize\csname
@twocolumnfalse\endcsname

\date{\today}
\maketitle

\begin{abstract}
The large $J_2$ limit of the square-lattice $J_1-J_2$ Heisenberg antiferromagnet
is a classic example of order by disorder where quantum fluctuations select
a collinear ground state. Here, we use series expansion methods
and a meanfield spin-wave theory to study the excitation
spectra in this phase and look for a finite
temperature Ising-like transition, corresponding to a broken symmetry of
the square-lattice, as first proposed by Chandra {\it et al.} (Phys. Rev. Lett.
{\bf 64}, 88 (1990)). We find that the spectra reveal the symmetries of
the ordered phase. However, we do not find any evidence for a finite
temperature phase transition. Based on an effective field theory we argue
that the  Ising-like transition occurs
only at zero temperature.
\end{abstract}


]

\narrowtext

The square-lattice $J_1-J_2$ Heisenberg antiferromagnet
is described by the Hamiltonian
\begin{equation}
H = J_1 \sum_{{\rm n.n.}} {\bf S}_i\cdot {\bf S}_j  +
 J_2 \sum_{{\rm n.n.n.}} {\bf S}_i\cdot {\bf S}_j \label{H}
\end{equation}
where the first sum runs over the nearest neighbor and the second over the
second nearest neighbor spin pairs of the square-lattice. For $J_1=0$,
the two sublattices are disconnected and
individually order antiferromagnetically.
For non-zero $J_1/J_2$ (less than some critical value),
in the classical ground state, the two sublattices
remain free to rotate with respect to each other. However, quantum
fluctuations lift this degeneracy
and select a collinear ordered state, where the neighboring spins
align ferromagnetically along one axis of the square-lattice and
antiferromagnetically along the other
\cite{Shender,chandra,chubukov,starykh}.
Thus the ground state breaks
both spin-rotational symmetry as well as the four-fold
symmetry of the square-lattice.

It is well known from the Mermin-Wagner
theorem that the continuous spin-rotational symmetry is restored in 2D at
any finite temperature. However, the discrete broken lattice symmetry
can in principle survive at finite temperatures. In an important paper, Chandra,
Coleman and Larkin \cite{chandra} argued that this symmetry should
be restored at a finite
temperature phase transition, which lies in the 2D Ising universality class,
and gave estimates for the transition temperature. They made use of a
linear spin-wave theory (LSWT) for the excitation spectra, which had gapless excitations
at four points of the Brillouin zone (($0,0$), ($\pi$,0), (0,$\pi$) and
($\pi,\pi$)).


Recent interest in this model comes from the discovery of two materials
\livo and \ligo by Melzi {\it et al.} \cite{melzi00,melzi01}. Electronic
structure calculations for these materials \cite{rosner02,rosner03}
lead to $J_2$ bigger than $J_1$,
perhaps by as much as an order of magnitude. The
reason for the unusually large $J_2$ can be qualitatively understood
from the crystal structure. The $VO_5$ pyramids alternately point up
and down, and hence, the spin-half vanadium atoms are alternately
displaced slightly above and slightly below the plane.
This causes an increased overlap with the second neighbors which fall in the
same plane. Many experimental features of the material are well described
by the $J_1-J_2$ Heisenberg model. However, the value of the ratio $J_1/J_2$
remains ill-determined \cite{misguich}.

It was argued by Rosner {\it et al.} \cite{rosner03} that the measurement
of the spin-wave spectra maybe particularly useful for establishing
this ratio. One of the purposes of this paper is to present quantitatively
accurate spectra for the model going beyond linear spin-wave theory.
Indeed, as we will show, the full spectra have gapless excitations
at only two symmetry related points of the Brillouin zone (0,0) and
(0,$\pi$), whereas the accidental degeneracies at ($\pi,\pi$) and ($\pi,0$)
are lifted by the order by disorder effect \cite{Shender}
(here we choose the x axis along the direction of the collinear
spin ordering, see Fig. 1).
Furthermore, the ratio
of the spin-wave velocities along x and y directions depends sharply
on the $J_1/J_2$ ratio and its measurement can
indeed be used to determine the latter ratio.

The real materials also have weak interplanar couplings, which eventually
lead to 3D long-range order and a finite temperature transition.
The specific heat data above the 3D transition does not have any clear
features which could be interpreted as a 2D Ising transition.
Indeed, the issue of
finite-temperature Ising-like
transitions may only be relevant to materials with sufficiently weak
interplanar couplings.
The issue is  conceptually important, however.
Can long-range Ising order, corresponding to antiferromagnetic correlations
along one axis and ferromagnetic correlations along the other,
survive at finite temperatures even after
the underlying spin-spin correlations become short ranged?
That is the somewhat paradoxical prediction of Chandra {\it et al.}\cite{chandra}

We calculate
the susceptibility appropriate for this Ising order by high temperature
series expansions. The numerical series extrapolations  fail to substantiate
a finite temperature Ising-like transition, they are only consistent
with a $T=0$ phase-transition in a strictly 2D model.
We also present field-theoretical arguments in favor of a $T=0$
phase-transition, and find the critical index for the
$T=0$ divergence of the susceptibility.

We begin by calculations of the excitation spectra
at zero temperature.
The standard linear spin-wave theory
predicts the dispersion relation\cite{chandra},
\be
\Delta = 2 J_2 \left[ \left( 1 + {J_1\over 2 J_2} \cos k_x \right)^2
 - \cos^2 k_y \left( \cos k_x + {J_1\over 2 J_2} \right)^2 \right]^{1/2}
\ee
and this results in a spectrum with zero modes at four points
${\bf k} =(0,0)$, $(0,\pi)$, $(\pi,\pi)$ and $(\pi,0)$,
as discussed by Chandra {\it et al.}\cite{chandra}.

We have derived the spin-wave dispersion using a meanfield spin-wave theory
(MFSWT), see e.g. Refs. \cite{starykh,Dots}. The result reads
\bea
\Delta &=& 2 J_2 \mu {\Big [} \left( 1 + y - \frac{x}{2} (1-\cos k_x) \right)^2 \nonumber \\
&& - \cos^2 k_y \left( \cos k_x + y \right)^2 {\Big ]}^{1/2}
\eea
where $\mu$, $x$ and $y$ are functions of $J_2/J_1$. For example for
\bea
&J_2/J_1=2:~~~ & \mu = 1.153,~ x=0.327, ~ y=0.174 \nonumber \\
&J_2/J_1=1:~~~ & \mu = 1.122,~ x=0.783, ~ y=0.452 \\
&J_2/J_1=0.6:~ & \mu = 0.818,~ x=1.583, ~ y=1.323 \nonumber
\eea
This dispersion give a spectrum with zero modes at only two points,
i.e. ${\bf k}=(0,0)$ and $(0,\pi)$.

We have performed an Ising  expansion\cite{oit96} of the excitation spectra
using the
linked-cluster series expansion method, as reviewed recently
by Gelfand and Singh \cite{gel00}.
The spin triplet excitation energy
$\Delta (k_x ,k_y)$ has been computed up to order $9$ for
the system in the collinear ordered phase.
A list of 1796200 linked clusters of up to 10 sites contribute to the
triplet excitation spectrum. The series are available on request.
Note that the dispersion relation has the
symmetry:
$\Delta (k_x ,k_y)=\Delta (k_x ,\pi-k_y)$.

\begin{figure}[htb]
\begin{center}
 \epsfig{file=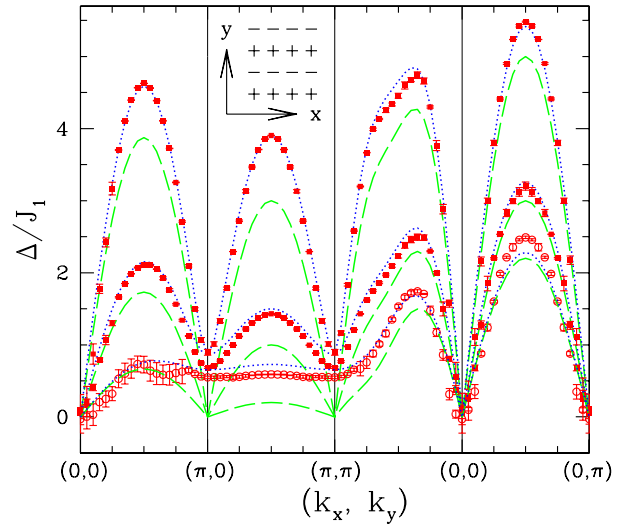,width=8.2cm}
  \vskip 5mm
 \caption[]
         {
Plot of the spin-triplet excitation spectrum
$\Delta (k_x, k_y)/J_1$ along high-symmetry cuts through the Brillouin
zone for the system with coupling ratios $J_2/J_1=2, 1, 0.6$
shown in the figure from the top to the bottom, respectively;
the blue dotted lines are the results of MFSWT,
the green dashed lines are the results of LSWT,
while the red points with  error bar are the results of series expansions.
The unperturbed spin configuration assumed in x and y directions
are shown as inset.
}
\label{fig_1}
\end{center}
\end{figure}

The excitation spectra for coupling ratios $J_2/J_1=2, 1, 0.6$ are shown in
Fig. \ref{fig_1}, together with those obtained from standard linear spin-wave
theory\cite{chandra} and
a MFSWT \cite{Dots}.
We can see from this figure that the spectrum does not
have the $90^{\rm o}$  rotational symmetry of the square-lattice.
There are zero modes at only two points i.e.
 ${\bf k}=(0,0)$ and $(0,\pi)$, as predicted by the
 MFSWT theory, but not the standard theory.
 In fact, the series expansion data agree with the predictions of the
 MFSWT theory with remarkable accuracy over the entire
 range of momenta, except for a small dip in the third panel.
The gaps at ${\bf k}=(\pi,\pi)$ and $(\pi,0)$ are
non-zero,  due to the
``order by disorder" effect\cite{Shender}.

The gap at ${\bf k}=(\pi,\pi)$ and the spin-wave velocity along $x$ and $y$-directions
versus $J_1/J_2$ are shown in Fig. 2 and Fig. 3. Clearly, the latter anisotropy
provides a good way to determine the $J_1/J_2$ ratio.

\begin{figure}[htb]
\begin{center}
 \epsfig{file=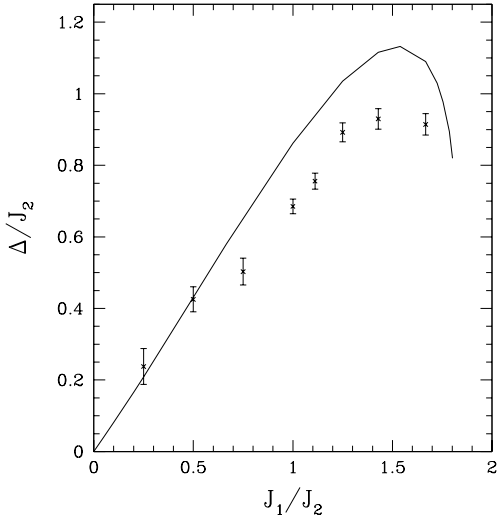,width=7cm}
  \vskip 5mm
 \caption[]
         {
The spin-triplet excitation gap $\Delta/J_2$ at ${\bf k}=(\pi,\pi)$
versus $J_1/J_2$ obtained from series expansion (the points), and
the MFSWT (the solid line).
}
 \label{fig_2}
 \end{center}
\end{figure}

\begin{figure}[htb]
\begin{center}
 \epsfig{file=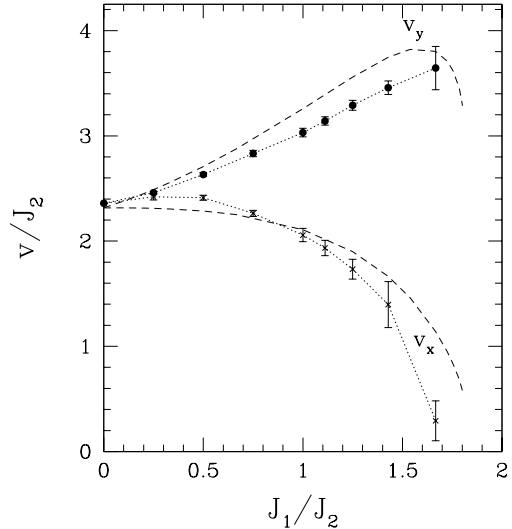,width=7cm}
  \vskip 5mm
 \caption[]
         {
The spin-wave velocity $v/J_2$ along x and y-directions
versus $J_1/J_2$ obtained from series expansion (the points with dotted lines connecting them), and
the MFSWT  (the dashed line).
}
 \label{fig_3}
 \end{center}
\end{figure}

We now turn to the question of an Ising-like finite temperature transition.
To explore the possibility of such a transition we calculate
the high temperature
series  for the susceptibility with respect to the field
\be
\label{f}
F = \sum_i ( {\bf S}_i \cdot {\bf S}_{i+\hat{x}} - {\bf S}_i \cdot {\bf S}_{i+\hat{y}} )
\ee
i.e. we compute the series for
\be
T\chi = \langle F^2 \rangle - \langle F \rangle^2 \label{eqchi}
\ee
where
\be
\langle Q \rangle = {{\rm Tr} Q \exp(-\beta H) \over {\rm Tr} \exp(-\beta H) }
\ee
and $\beta=1/(k_B T)$; we take $k_B=1$  here.
Note that $\langle F \rangle$ is zero for the bulk system, but
we need
to include it in Eq. (\ref{eqchi}) because it is not zero
for each individual cluster in the series
expansion.

This field breaks the 90 degree rotational symmetry, as  in the zero temperature
collinear ordered phase, and if there was a finite temperature transition, one
would expect $\chi$ to diverge at the critical temperature.
The series has been computed up to order $\beta^8$,
using the same list of 1796200 linked
clusters of up to 10 sites as  for the spin dispersion.
The series are available on request.

\begin{figure}[htb]
\begin{center}
 \epsfig{file=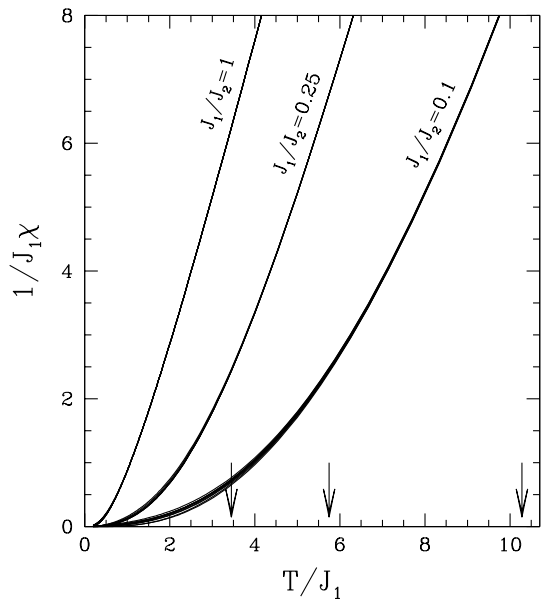,width=8cm}
  \vskip 5mm
 \caption[]
         {
The inverse of susceptibility $1/J_1\chi$  versus $T/J_1$ for
$J_1/J_2=1, 0.25, 0.1$.
Several different integrated differential
approximants to the high-temperature series are shown.
The Ising-like transition temperatures estimated
according to \cite{chandra} are shown by arrows.
}
 \label{fig_4}
 \end{center}
\end{figure}

For the full series, we first tried to locate the critical point by Dlog
Pad\'e approximants\cite{gut}, but this did not give
any consistent results.
We also used integrated differential approximants\cite{gut}
to extrapolate the series, and the results are shown in Fig. \ref{fig_4}.
We can see that $1/\chi$ appears to vanish only at zero temperature, rather than
at finite temperature. This is consistent also with the results of the
Pad\'e approximants. The Ising-like transition temperatures estimated
according to \cite{chandra} are shown in  Fig. \ref{fig_4} by arrows.

To look further into this issue, we have also re-analyzed our high-temperature
series for the specific heat $C_v$\cite{rosner02,rosner03},
for the ratio of couplings in the collinear ordered phase. While the
convergence of the series extrapolation is not very good at low temperatures,
the Dlog Pad\'e approximants to the series  for $C_v$ do not show evidence for a
finite temperature critical point \cite{misguich}.

We also analyse our numerical results using an effective field theory.
The dynamical variables of the problem are two vector fields ${\bf n}_1$ and ${\bf n}_2$
corresponding to two interpenetrating Neel sublattices with exchange
interaction $J_2$. Each of the fields is described by the nonlinear
$\sigma$-model, and there is also an interaction between the fields.
The effective Lagrangian of the system reads \cite{chandra}
\be
\label{le}
L_{\rm eff}= \frac{1}{2}\int d^2 x\left[\sum_{i=1,2}(\partial_{\mu}{\bf n}_i)^2
  -g({\bf n}_1\cdot{\bf n}_2)^2\right] \ ,
\ee
where $g \sim J_1^2/J_2$ is the coupling constant. There are two degenerate
energy minima corresponding to $\sigma={\bf n}_1\cdot{\bf n}_2=\pm 1$. So
$\sigma$ imitates an Ising parameter, and at $T=0$ it takes  a values
corresponding to one of the minima,
$\sigma =1$ (x-collinear state) or $\sigma =-1$ (y-collinear state).
However, we stress that there is not a dynamical Ising variable in the
system, the entire dynamics being described in terms of ${\bf n}_1$ and ${\bf n}_2$
and the Lagrangian (\ref{le}). The two minima $\sigma=\pm 1$ are separated by
a potential barrier $W(0) \sim EL^2$, where $E\sim g$ is  the
interaction energy density, and $L \to \infty$ is the size of the system.

At a finite temperature, each sublattice behaves as an
ordered Neel state up to a length scale $\xi \approx
 0.303 \exp(2\pi\rho_s/T) [1 - T/(4 \pi \rho_s) ]$, where $\rho_s \approx
 0.182J_2$ is the spin stiffness \cite{CHN}.
At low temperature the energy barrier required to move from the $\sigma =+1$ configuration to
the $\sigma=-1$ configuration, $W(T) \sim E \xi^2$, remains finite at any finite $T$!
So, using the quantum field theory language, one has to say that there is a
nontrivial instanton in the problem with tunneling  probability $\propto \exp(-E\xi^2/T)$.
This is different from the true Ising situation, where there is a {\it dynamical}
Ising field with stiffness, and the tunneling probability $\propto \exp(-E(T)
L^2/T) \to 0$, as $L \to \infty$. The ``true" Ising transition occurs when $E(T)$ vanishes.

\begin{figure}[htb]
\begin{center}
 \epsfig{file=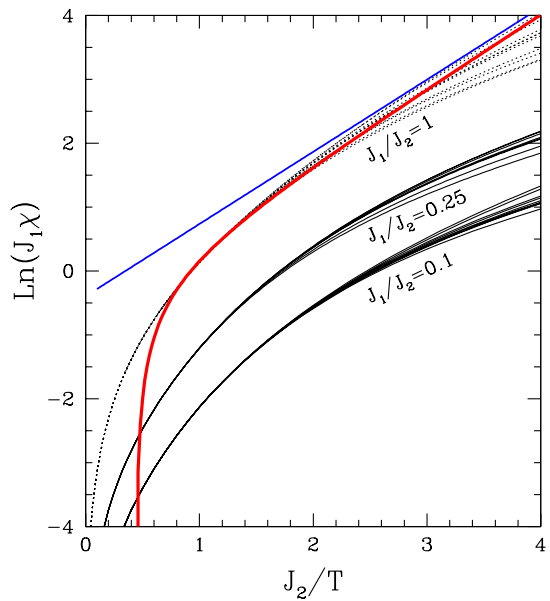,width=8cm}
  \vskip 5mm
 \caption[]
{ $\ln(J_1\chi)$  versus $J_2/T$ for $J_1/J_2=1, 0.25, 0.1$.
Also shown are $0.8+\ln\xi$ with (red bold line) and
without (blue straight line) the
prefactor $[1-T/(4\pi\rho_s)]$.
}
 \label{fig_5}
 \end{center}
\end{figure}

Based on the picture described we immediately come to the conclusion that at
any finite $T$ the system fluctuates beween $\sigma =1$ and $\sigma =-1$ local
minima, and hence the transition temperature for the Ising-like transition is
zero in agreement with numerical data. According to this picture one would
expect that at low temperature and at $J_2> J_1$ the suceptibilty with respect
to the field (\ref{f}) behaves as
\be
\label{che}
\chi \propto \xi^{\alpha} \propto e^{-1.15\alpha J_2/T} \ ,
 \ee
where $\alpha$ is a critical index,
(we neglect the prefactor $[1-T/(4\pi\rho_s)]$ in $\xi$).
Plots of $\ln(J_1\chi)$ versus $J_2/T$ are
presented in Fig.  \ref{fig_5}.
The plots  are roughly consistent with prediction (\ref{che})
with the critical index $\alpha \approx 1$.

In conclusion, in this paper we have studied the collinear phase
of the $J_1-J_2$ square-lattice Heisenberg antiferromagnet by series
expansion methods. We have obtained quantitatively accurate excitation spectra,
which are significantly more accurate than the spin-wave calculations\cite{chandra},
and should be helpful in determining the exchange
parameters for materials with large second neighbor interactions.
The MFSWT predicts these spectra with great accuracy.
We have also studied an Ising-like phase transition in this model.
Using high temperature expansions for the appropriate susceptibility
as well as quntum field theory arguments we have shown that the
transition temperature is exactly zero.

This work is supported by a grant
from the Australian Research Council and by US National Science Foundation
grant number DMR-9986948.
The computation has been performed on an AlphaServer SC computer.
We are grateful for the computing resources provided
by the Australian Partnership for Advanced Computing (APAC)
National Facility.

\bibliography{basename of .bib file}


\end{document}